\documentclass[reprint,amsmath,amsfonts,amssymb,aps,prx,preprintnumbers,superscriptaddress]{revtex4-2}

\usepackage[utf8]{inputenc}
\usepackage[T1]{fontenc}
\usepackage{lmodern}
\usepackage{graphicx}
\usepackage{dcolumn}
\usepackage{amsmath}
\usepackage{bm}
\usepackage{textcomp}
\usepackage{ulem}
\usepackage{ifpdf}
\usepackage[squaren,Gray]{SIunits}
\usepackage{color}
\definecolor{red}{rgb}{1,0,0}
\definecolor{blue}{rgb}{0,0,1}
\definecolor{darkred}{rgb}{0.6,0,0}
\definecolor{darkblue}{rgb}{0,0,.6}
\definecolor{darkgreen}{rgb}{0,0.5,0}

\usepackage{ifpdf}
\ifpdf
\usepackage{epstopdf}
\usepackage[pdftex,unicode,pdfstartview={FitH},pdfborder={0 0 0}]{hyperref}
\usepackage{hypcap}
\else
\usepackage[hypertex]{hyperref}
\fi
\hypersetup{
    bookmarksnumbered = true,
    colorlinks = true, linkcolor = black,
    citecolor = black, filecolor = black,
    menucolor = black, urlcolor = black
}

\newcolumntype{R}{>{$\displaystyle}r<{$}}
\newcolumntype{C}{>{$\displaystyle}c<{$}}

\begin{document}
    
    \title{Hybrid moir\'e excitons and trions in twisted MoTe$_2$-MoSe$_2$ heterobilayers}
    
     \author{Shen Zhao}
    \affiliation{Fakult\"at f\"ur Physik, Munich Quantum Center, and Center for NanoScience (CeNS), Ludwig-Maximilians-Universit\"at M\"unchen, Geschwister-Scholl-Platz 1, 80539 M\"unchen, Germany}

    \author{Xin Huang}
    \affiliation{Beijing National Laboratory for Condensed Matter Physics, Institute of Physics, Chinese Academy of Sciences and School of Physical Sciences, CAS Key Laboratory of Vacuum Physics, University of Chinese Academy of Sciences, Beijing 100190, P. R. China}
    \affiliation{School of Physical Sciences, CAS Key Laboratory of Vacuum Physics, University of Chinese Academy of Sciences, Beijing 100190, P. R. China}    
        \affiliation{Fakult\"at f\"ur Physik, Munich Quantum Center, and Center for NanoScience (CeNS), Ludwig-Maximilians-Universit\"at M\"unchen, Geschwister-Scholl-Platz 1, 80539 M\"unchen, Germany}
    
    \author{Roland Gillen}
    \affiliation{Department of Physics, Friedrich-Alexander Universit\"at Erlangen-N\"urnberg, Staudtstra{\ss}e 7, 91058 Erlangen, Germany}
    
    \author{Zhijie Li}
    \affiliation{Fakult\"at f\"ur Physik, Munich Quantum Center, and Center for NanoScience (CeNS), Ludwig-Maximilians-Universit\"at M\"unchen, Geschwister-Scholl-Platz 1, 80539 M\"unchen, Germany}
    
    \author{Song Liu}
    \affiliation{Department of Mechanical Engineering, Columbia University, New York, New York 10027, United States}
    
    \author{Kenji Watanabe}
    \affiliation{Research Center for Functional Materials, National Institute for Materials Science, 1-1 Namiki, Tsukuba 305-0044, Japan}
    
    \author{Takashi Taniguchi}
    \affiliation{International Center for Materials Nanoarchitectonics, National Institute for Materials Science, 1-1 Namiki, Tsukuba 305-0044, Japan}
    
    \author{Janina Maultzsch}
    \affiliation{Department of Physics, Friedrich-Alexander Universit\"at Erlangen-N\"urnberg, Staudtstra{\ss}e 7, 91058 Erlangen, Germany}

    \author{James Hone}
    \affiliation{Department of Mechanical Engineering, Columbia University, New York, New York 10027, United States}    
    
    \author{Alexander H{\"o}gele}
    \affiliation{Fakult\"at f\"ur Physik, Munich Quantum Center, and Center for NanoScience (CeNS), Ludwig-Maximilians-Universit\"at M\"unchen, Geschwister-Scholl-Platz 1, 80539 M\"unchen, Germany}
    \affiliation{Munich Center for Quantum Science and Technology (MCQST), Schellingstra\ss{}e 4, 80799 M\"unchen, Germany}

\author{Anvar~S.~Baimuratov}
    \affiliation{Fakult\"at f\"ur Physik, Munich Quantum Center, and Center for NanoScience (CeNS), Ludwig-Maximilians-Universit\"at M\"unchen, Geschwister-Scholl-Platz 1, 80539 M\"unchen, Germany}
    \date{\today}

\begin{abstract}
We report experimental and theoretical studies of MoTe$_2$-MoSe$_2$ heterobilayers with rigid moir\'e superlattices controlled by the twist angle. Using an effective continuum model that combines resonant interlayer electron tunneling with stacking-dependent moir\'e potentials, we identify the nature of moir\'e excitons and the dependence of their energies, oscillator strengths and Land\'e $g$-factors on the twist angle. Within the same framework, we interpret distinct signatures of bound complexes among electrons and moir\'e excitons in nearly collinear heterostacks. Our work provides fundamental understanding of hybrid moir\'e excitons and trions in MoTe$_2$-MoSe$_2$ heterobilayers, and establishes the material system as a prime candidate for optical studies of correlated phenomena in moir\'e lattices.
\end{abstract}

\maketitle

In semiconducting transition metal dichalcogenide (TMD) heterostructures, electrons and holes located in distinct layers are bound by strong Coulomb attraction, forming interlayer excitons~\cite{Rivera2018}. In contrast to intralayer excitons confined within one TMD layer, interlayer excitons are much longer-lived, exhibiting radiative lifetimes in the nanosecond range and facilitating long-range exciton- and valley-spin transport~\cite{Jin2018,Unuchek2018}. In addition, the permanent out-of-plane electric dipole enables wide spectral tunability through the quantum-confined Stark effect~\cite{Jauregui2019,Ciarrocchi2019}. More recently, interlayer excitons in TMD moir\'e heterostructures have been employed to probe correlated electronic states~\cite{Miao2021,Tan2023}, and have been observed to form bosonic correlated states~\cite{Park2023,lian2023}.

Due to their spatially indirect nature, however, interlayer excitons in TMD heterostructures have more reduced electron-hole wave function overlap~\cite{Gillen2018}, which results in a negligible oscillator strength and thus substantially limits their coupling to light. One strategy to increase light-matter coupling of interlayer excitons relies on admixing of bright intralayer excitons through tunneling of electrons or holes in near-resonant moir\'e homobilayers and heterostructures. Related examples include MoS$_2$ and WSe$_2$ homobilayers~\cite{Leisgang2020,Tagarelli2023} as well as MoSe$_2$-WS$_2$ \cite{Tang2021,Polovnikov2024} and MoSe$_2$-hBN-MoSe$_2$ \cite{Shimazaki2020} heterostructures. In MoSe$_2$-WS$_2$ heterobilayers, in particular, near-resonant conduction bands enhance the hybridization among interlayer and intralayer excitons and give rise to a multiplicity of moir\'e excitons with sizable oscillator strength~\cite{Polovnikov2024}. This, in turn, provides all-optical means to probe the physics of correlated charges in moir\'e lattices~\cite{Shimazaki2020,Ciorciaro2023}.

\begin{figure*}[t!]
    \includegraphics[scale=1.05]{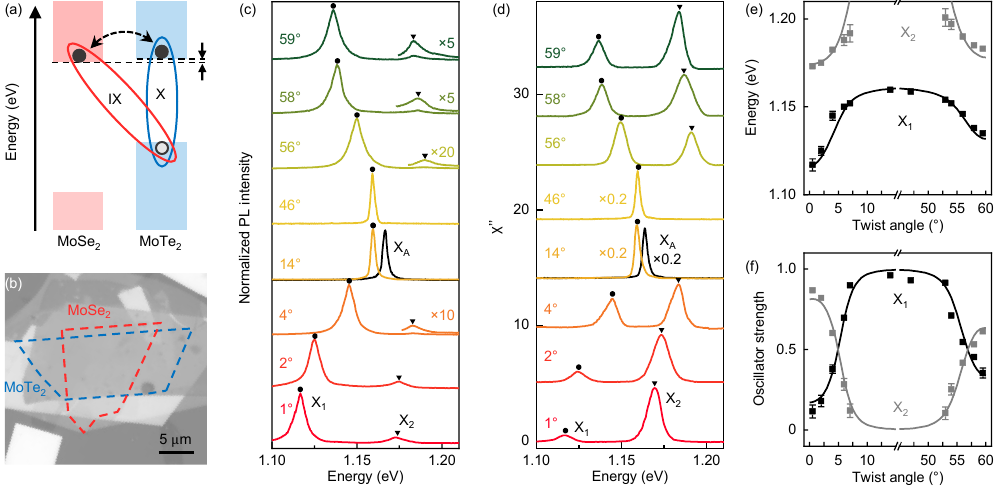}
    \caption{(a) Schematic band diagram of MoTe$_2$-MoSe$_2$ in type-II band alignment with near-degenerate conduction band edges. The dashed arrow indicates resonant interlayer electron tunneling, which promotes the hybridization between interlayer exciton IX and intralayer exciton X states of MoTe$_2$. (b) Optical micrograph of a representative hBN-encapsulated heterobilayer with a twist angle of $58^\circ$. The edges of monolayer MoTe$_2$ and MoSe$_2$ are marked with dashed lines. (c) and (d) Normalized PL and absorption spectra (calculated as imaginary susceptibility~$\chi''$ from differential reflectivity via Kramers-Kronig transform), respectively, for heterobilayers with various twist angles at charge neutrality. Two characteristic peaks, X$_1$ and X$_2$, are highlighted by black dots and triangles, respectively. In both panels, single-peak spectra of monolayer MoTe$_2$ exciton X$_\text{A}$ are shown in black lines. (e) and (f) Evolution of energy and oscillator strength for the peaks X$_1$ and X$_2$ as a function of the twist angle, respectively. The oscillator strength was obtained as the integrated peak area in absorption normalized to the peak area of X$_\text{A}$ in monolayer MoTe$_2$. In both panels, the results of an effective moir\'e model (solid lines) fitted to the data (squares) yield the hopping parameter $t = 12$~meV and the energy difference between the decoupled X and IX exciton states as $54$~meV, consistent with the \textit{ab initio} conduction band offset of $38$~meV~\cite{Zhang2016}.}
    \label{fig1}
\end{figure*}

In this work, we demonstrate that MoTe$_2$-MoSe$_2$ heterobilayers realize canonical moir\'e systems with near-resonant conduction bands. The large mismatch of $\sim7\%$ between the lattice constants of MoTe$_2$ and MoSe$_2$ layers renders the system robust against mesoscopic lattice reconstruction~\cite{Zhao2023}, thus supporting moir\'e interference phenomena down to collinear alignment in parallel (R-type) and antiparallel (H-type) limits. The theoretically predicted conduction band offset of 38~meV is very small as compared to hundreds of meV in MoSe$_2$-WSe$_2$ and MoS$_2$-WSe$_2$ heterostacks~\cite{Zhang2016}. We demonstrate that this small offset promotes sizable hybridization between interlayer (IX) and intralayer excitons (X)  that share the hole state in the MoTe$_2$ layer and form moir\'e excitons~\cite{Deilmann2018} with high sensitivity to the twist angle of the MoTe$_2$-MoSe$_2$ heterostack. Moreover, we provide a theoretical framework that captures the main optical signatures of moir\'e excitons and their charged counterparts induced by field-effect doping.

The samples based on MoTe$_2$-MoSe$_2$ heterobilayers with different twist angles~$\theta$ were stacked from exfoliated monolayers and encapsulated in thin hexagonal boron nitride (hBN) to ensure spectrally narrow exciton resonances~\cite{Cadiz2017}. For field-effect doping, the heterostacks were connected to a charge reservoir, and a few-layer graphite bottom gate was used to control the doping level via the gate voltage. Details of sample fabrication and determination of the twist angle $\theta$ are provided in Supplementary Note~1 and 4. Figure~\ref{fig1}(b) shows the optical micrograph of a representative sample, with very clean interfaces between different layers. Owing to the large lattice mismatch, the heterostacks are rigidly locked in the canonical moir\'e geometry. Consistently, the spectral features in absorption and photoluminescence (PL) are very homogeneous across each MoTe$_2$-MoSe$_2$ sample (see Supplementary Note 5) without sizable variations.

\begin{figure*}[t!]
    \includegraphics[scale=1.05]{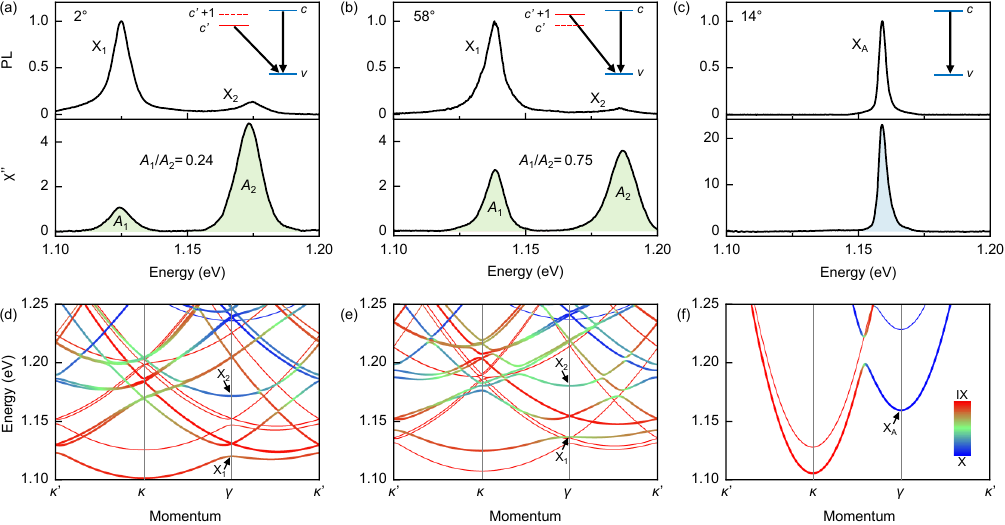}
    \caption{(a--c) PL (upper panels) and absorption (lower panels) spectra of heterobilayers with $2^\circ$ (a), $58^\circ$ (b) and $14^\circ$ (c) twist angles. Note that the absorption strength ratio ($A_1/A_2$) between $X_1$ and $X_2$ in stacks with $2^\circ$ and $58^\circ$ twist is very different despite the same rotation angle away from collinear alignment. For each twist angle, the inset illustrates the spin-polarized bands of MoSe$_2$ (red) and MoTe$_2$ (blue), with solid and dashed lines represent spin-up and spin-down polarized bands and the arrows indicate optical transitions involving spin-singlet excitons. (d--f) Calculated exciton dispersions in the mini Brillouin zone for twisted heterobilayers with twist angles as in (a--c). The color code in the inset of (f) represents the state composition, ranging from pure intralayer exciton X (blue) to pure interlayer exciton IX (red) character. Thick and thin lines represent singlet and triplet branches, and the arrows in each panel indicate the moir\'e states $X_1$ and $X_2$ with zero center-of-mass momentum involved in optical transitions.}
    \label{fig2}
\end{figure*}

Representative cryogenic PL and absorption spectra of MoTe$_2$-MoSe$_2$ heterobilayers at 3.2~K are shown in Fig.~\ref{fig1}(c) and (d). For heterobilayers close to parallel R-type alignment ($\theta=1^\circ$, $2^\circ$ and $4^\circ$) and antiparallel H-type alignment ($\theta=56^\circ$, $58^\circ$ and $59^\circ$), the PL spectra feature two peaks, X$_1$ and X$_2$, located respectively at the low- and high-energy side of the A-exciton resonance of monolayer MoTe$_2$ at 1.168~eV labelled as X$_\text{A}$~\cite{Ruppert2014,Lezama2015,Froehlicher2016,Robert2016}. As shown in Fig.~\ref{fig1}d, both peaks contribute to absorption with a spectral weight that depends on the twist angle. The peaks arise from angle-dependent hybridization of pure inter- and intralayer states, with admixed moir\'e excitons acquiring their oscillator strength from the fundamental exciton X$_\text{A}$ of monolayer MoTe$_2$. Contrastingly, for the largely misaligned heterostacks ($\theta=14^\circ$ and $46^\circ$), the PL and absorption spectra exhibit only one single peak with oscillator strength and energy similar to the monolayer X$_\text{A}$ exciton. This is consistent with a fully decoupled intralayer exciton in the heterostack, where the small energy red-shift from the X$_\text{A}$ transitions is due to Coulomb screening by the adjacent MoSe$_2$ layer~\cite{Raja2017}.
In addition to the stark differences between nearly aligned and largely twisted heterobilyers, the spectral features exhibit a continuous evolution with the twist angle, as evident from the data in Fig.~\ref{fig1}(e) and (f). The trend is similar in R- and H-type stacks: both X$_1$ and X$_2$ feature peak energy minima in Fig.~\ref{fig1}(e) for near-collinear alignments at $\theta\approx0^\circ$ and $60^\circ$, where the absorption strength in Fig.~\ref{fig1}(f) is minimal for X$_1$ and maximal for X$_2$. For increasing twist angles, the peak energies of both states increase continuously, and the oscillator strength is gradually transferred from X$_2$ to X$_1$. For angles above $7^\circ$ (below $53^\circ$), the contribution of X$_2$ to both PL and absorption vanishes, whereas X$_1$ evolves into a state of pure intralayer character with almost constant absorption strength and transition energy for $7^\circ<\theta<53^\circ$. Note that this evolution of the oscillator strength with the twist angle also manifests in the exciton recombination dynamics, with the lifetime of X$_1$ reducing continuously from several hundreds of ps in nearly collinear heterostacks to less than $10$~ps in largely twisted ones ($7^\circ<\theta<53^\circ$), shown in Supplementary Note~6.

Despite the overall similar trends, it is instructive to note that R- and H-type stacks exhibit important dissimilarities when the angle deviation from ideal alignment is the same. As exemplified in Fig.~\ref{fig2}(a) and (b) for heterostacks with $2^\circ$ and $58^\circ$ twist, respectively, the moir\'e peak doublets are red-shifted by $\sim 13$~meV with respect to each other, with X$_1$ and X$_2$  at $1.125$ and $1.175$~eV in Fig.~\ref{fig2}(a) and $1.138$ and $1.187$~eV in Fig.~\ref{fig2}(b). Moreover, the absorption strength ratio ($A_1/A_2$) of the moir\'e peaks X$_1$ and X$_2$ is about three times smaller for near-parallel alignment ($A_1/A_2=0.24$) as compared to the near-antiparallel alignment ($A_1/A_2=0.75$). Such differences in peak energies and absorption strength ratios are also evident in Fig.~\ref{fig1}(e) and (f) for stacks with $1^\circ$ and $59^\circ$ as well as $4^\circ$ and $56^\circ$ twist angles.

\begin{figure*}[t!]
    \includegraphics[scale=1.05]{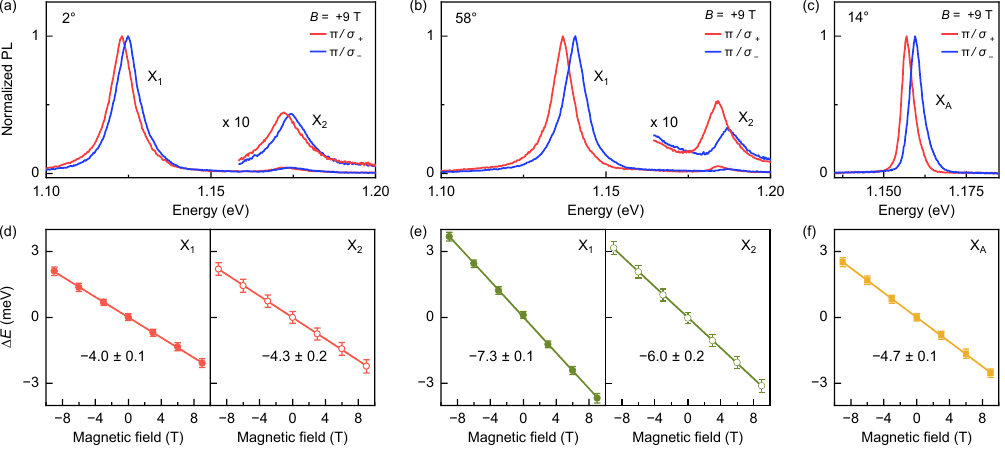}
    \caption{Magneto-luminescence spectroscopy of heterobilayers with $2^\circ$, $58^\circ$ and $14^\circ$ twist angles. (a--c) PL spectra in perpendicular magnetic field of $9$~T recorded with linearly polarized excitation ($\pi$) and circularly polarized detection ($\sigma_+$ and $\sigma_-$). For energies above $1.16$~eV, spectra scaled by a factor of $10$ in intensity were also included in (a) and (b) for clarity. (d--f) Zeeman splitting $\Delta E$ of the exciton peaks in (a--c) as a function of the magnetic field. The error bars originate from Voigt fits to the spectra. The respective exciton $g$-factors with least-square errors are extracted from linear fits (solid lines) and explicitly given in each panel.}
    \label{fig3}
\end{figure*}

To explain this set of spectral features and their evolution with the twist angle, we employed the phenomenological model of near-resonant hybridization in moir\'e heterostructures \cite{Polovnikov2024} (see Supplementary Note 8 for details). In brief, the model has two main contributions. First, the real-space moir\'e pattern is accounted for in the framework of the continuum model~\cite{WuTopo2017}, where harmonic expansion yields a complex moir\'e potential $V=|V| \exp(i\phi)$ with depth $|V|$ and phase $\phi$ characteristic of a given heterostructure and stacking type. Second, interlayer hybridization gives rise to electron Bragg scattering and the formation of a mini Brillouin zone~\cite{Ruiz-Tijerina2019}, with high symmetry points $\gamma$ and $\kappa$ corresponding to the Brillouin zone corners of MoTe$_2$ and MoSe$_2$ monolayers, respectively. Interlayer tunnel coupling mixes the near-resonant conduction bands of MoTe$_2$ and MoSe$_2$ through electron hopping with parameter $t$, which gives rise to hybridization of intralayer and interlayer excitons X and IX with holes in the MoTe$_2$ layer and the respective energies $E_\text{X}=E_0$ and $E_\text{IX}=E_0+\Delta$, where $\Delta$ quantifies the difference between the energy of IX excitons and the energy $E_0$ of the fundamental A-exciton in MoTe$_2$. To a first approximation~\cite{Wang2017}, $t$ is treated as a constant inside the mini Brillouin zone, and only spin-conserving electron tunneling can take place, involving different spin-split conduction bands of MoSe$_2$ in the hybridization of R- and H-type limits. We also assume that the twist angle has a negligible impact on the interlayer distances, local atomic arrangement, hopping parameters and moir\'e potential amplitude.

The results of the combined effective model are shown as moir\'e exciton dispersions in Fig.~\ref{fig2}(d)--(f) for three different cases, corresponding to the heterostacks with PL and absorption spectra in Fig.~\ref{fig2}(a)--(c), respectively. Thick and thin lines represent spin singlet and triplet bands, and the arrows highlight the two optically active singlet moir\'e exciton states located at the $\gamma$ point of the mini Brillouin zone due to hole-localization in the MoTe$_2$ layer. These two states correspond to bright moir\'e excitons with zero center-of-mass momentum and dipole-allowed direct optical transitions. All other states at $\gamma$ are dark, as their admixing with the bright fundamental A-exciton of MoTe$_2$ is inhibited by symmetry. The color scale inset of Fig.~\ref{fig2}(f) indicates the relative contributions of X and IX excitons to the bands. 
    
The moir\'e exciton band structures in Fig.~\ref{fig2}(d)--(f) are shown for best-fit parameters $E_0 = 1162$~meV, $\Delta = -54$~meV, $t = 12$~meV, and $|V| = 1$~meV, with stacking-dependent phases $\phi = 3\pi/4$ and $7\pi/12$ for R- and H-type limits. The values were obtained from best correspondence between twist-dependent energies and oscillator strengths of moir\'e states X$_1$ and X$_2$ in our simulations and experiments shown by the solid lines in Fig.~\ref{fig1}(e) and (f) (see Supplementary Note 8 for details), taking into account the response of moir\'e excitons to electron doping discussed further below. Our model provides an intuition for the evolution of the energies and oscillator strength of both moir\'e peaks: the energy shifts of X$_1$ and X$_2$ stem from angle-dependent hybridization between X and IX excitons, which in turn redistributes the oscillator strength of the A-exciton in MoTe$_2$ among the hybrid moir\'e states. 

With this intuition, we understand the single optically active state in the spectra of largely twisted stacks as in Fig.~\ref{fig2}(c) as an almost pure A-exciton of MoTe$_2$, and the presence of two bright states in the spectra of nearly collinear stacks in Fig.~\ref{fig2}(a) and (b) with $\theta=2^\circ$ and $58^\circ$, respectively, as layer-hybridized moir\'e states with relatively flat dispersion and large effective masses. We also note that the admixing of IX and X exciton states is different for moir\'e excitons X$_1$ and X$_2$ for same angle deviations from ideal R- and H-alignments, e.g. as for $\theta = 2^\circ$ and $58^\circ$ in Fig.~\ref{fig2}. This difference is due to the reversed spin-ordering of decoupled IX bands in R- and H-type heterobilayers, rendering both the energy and the oscillator strength of hybrid moir\'e excitons asymmetric for angles close to $0^\circ$ and $60^\circ$, as in Fig.~\ref{fig1}(e) and (f).

Magneto-luminescence studies provide further quantitative support for our analysis. From magneto-dispersed polarization-contrasting PL spectra as in Fig.~\ref{fig3}(a)--(c) at $9$~T, recorded for heterobilayers with $\theta=2^\circ$, $58^\circ$, and $14^\circ$ twist angle, respectively, we deduce the characteristic $g$-factors of moir\'e exciton peaks. The corresponding Zeeman splittings $\Delta E$, given by the energy difference between the maxima of $\sigma_+$ and $\sigma_-$ polarized peaks, are shown in Fig.~\ref{fig3}(d)--(f) as a function of the magnetic field $B$ applied perpendicularly to the heterostack. Linear fits to the data with $\Delta E = g \mu_B B$ and the Bohr magneton $\mu_B$, yield the exciton $g$-factors of $-4.7\pm0.1$ for X$_A$ in the heterostructure with $14^\circ$ twist, as well as $-4.0\pm0.1$ and $-4.3\pm0.2$ and $-7.3\pm0.1$ and $-6.0\pm0.2$ for X$_1$ and X$_2$ moir\'e peaks in heterostacks with $2^\circ$ and $58^\circ$ twist angles, respectively. The $g$-factor of the single peak X$_A$ in the spectra of stacks with large twist angles is within error bars identical to the value of the fundamental exciton in monolayer MoTe$_2$, determined as $-4.6\pm0.1$ (see Supplementary Fig.~S1), whereas the $g$-factors of both moir\'e peaks in nearly collinear stacks are distinct.  

We employed \textit{ab initio} calculations for theoretical estimates of the exciton $g$-factor in monolayer MoTe$_2$ and moir\'e excitons in MoTe$_2$-MoSe$_2$ heterostructures~\cite{Foerste2020,Deilmann2020,Wozfiniak2020,Forg2021}. For the fundamental exciton X$_A$ in monolayer MoTe$_2$, we obtained $g_\text{A} = -4.6$ from: 
\begin{equation}
g_\text{A} = 2 (L_c - L_v),
\end{equation}
with the orbital angular momenta of the lowest conduction band and the highest valence band $L_c$ and $L_v$. For moir\'e excitons, we decompose the $g$-factors into the individual contributions of interlayer and intralayer excitons~\cite{Junior2023} according to hybridization-induced admixing as:
\begin{align}
    g_i^\text{(R)} &= 2 \left( f_i^\text{(X)} L_c  + f_i^\text{(IX)} L_{c'} - L_v \right), \\
    g_i^\text{(H)} &= 2 \left( f_i^\text{(X)} L_c  - f_i^\text{(IX)} L_{c'+1} - L_v \right),
\end{align}
where $i = 1,2$ indicates the hybrid X$_i$ exciton, $L_{c' (c'+1)}$ is the first (second) conduction band in MoSe$_2$ layer as shown in the insets of Fig.~\ref{fig2}(a) and (b), and  $f_i^\text{(X)}$ ($f_i^\text{(IX)}$) is the fraction of X (IX) exciton. Using the calculated values of the orbital angular momenta in Table S2 and the respective exciton fractions deduced from our effective model, we obtain $g_1^\text{(R)} = -4.2$ and $g_2^\text{(R)} = -4.4$ for X$_1$ and X$_2$ moir\'e excitons in heterostacks with $2^\circ$ twist, as well as $g_1^\text{(H)} = -8.0$ and $g_2^\text{(H)} = -6.8$ in stacks with $58^\circ$ twist angle. The overall agreement between the theoretical and experimental $g$-factor values provides strong confidence in our understanding of the intralayer and interlayer exciton character of moir\'e excitons X$_1$ and X$_2$. The small discrepancies between the theoretical and experimental values are not surprising given the finite extent of excitons in momentum space, with contributions of different orbital angular momentum values away from the $K$ point \cite{Polovnikov2024} unaccounted for in our model.

\begin{figure}[t!]
    \includegraphics[scale=1.05]{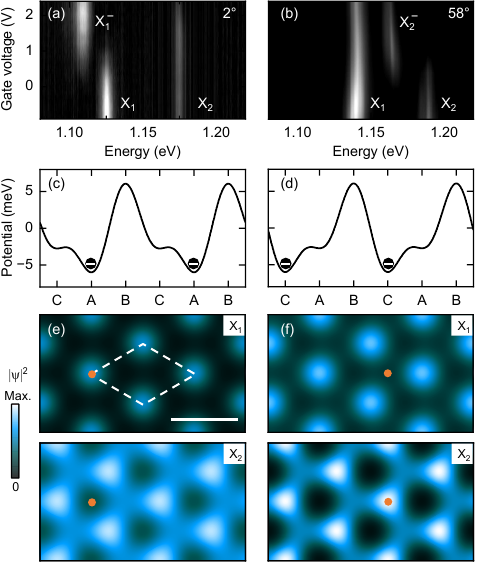}
    \caption{Hybrid moir\'e trions in MoTe$_2$-MoSe$_2$ heterostacks. (a) and (b) PL evolution with electron doping induced by the gate voltage in heterostacks with twist angles of $2^\circ$ and $58^\circ$, respectively. (c) and (d) Moir\'e potential for electrons in MoTe$_2$ across the high-symmetry stacking sites A, B, and C of the moir\'e unit cell shown in (e) by the dashed line (scale bar: 5~nm). (e) and (f) Color-coded maps of real-space probability functions $|\psi|^2$ of intralayer excitons contributing to X$_1$ (upper panel) and X$_2$ (lower panel) excitons. Orange dots indicate in each panel one representative electron-localizing site at low doping levels.}
    \label{fig4}
\end{figure}

Remarkably, the decomposition of moir\'e excitons into their respective intralayer and interlayer exciton fractions is the key to the understanding of the distinct behaviors observed upon electron doping in Fig.~\ref{fig4}(a) and (b) for stacks with $2^\circ$ and $58^\circ$ twist angles. In the PL of the stack close to R-type alignment in Fig.~\ref{fig4}(a), electron doping at elevated positive gate voltages induces a trion or attractive Fermi-polaron feature X$^-_1$ red-shifted by $14$~meV from the moir\'e peak X$_1$, whereas the peak X$_2$ remains almost unaffected. The PL of the stack near H-type alignment in Fig.~\ref{fig4}(b), on the other hand, exhibits the signature of the bound charge complex X$^-_2$ with a binding energy of $18$~meV, with X$_1$ being merely affected by doping. This contrasting response to electron doping in R- and H-type stacks, where at low doping electrons occupy the MoTe$_2$ layer first due to its lower conduction band edge, stems from both different hybridization characters of moir\'e excitons and distinct spatial positions of field-induced electrons inside the moir\'e unit cell~\cite{Jin2019,Naik2022,Polovnikov2024}.

To highlight these differences, we plot the moir\'e potential for electrons across the unit cell for nearly collinear heterostacks with $\theta = 2^\circ$ and $58^\circ$ in Fig.~\ref{fig4}~(c) and (d), respectively, and in Fig.~\ref{fig4}~(e) and (f) the spatial distributions of intralayer exciton wavefunctions contributing to X$_1$ (top panels) and X$_2$ (bottom panels) moir\'e excitons. Taking into account that electron-hole interactions are larger for intralayer charges than in their layer-separated limit~\cite{Liu2021}, the localization of the hole in the MoTe$_2$ monolayer favors the formation of bound exciton-charge complexes for electrons that are spatially co-localized with the interlayer part of moir\'e excitons. In moir\'e unit cells of R-type heterostacks, this is the case at sites A, where the spatial contribution of the intralayer exciton wavefunction is maximal for X$_1$ and minimal for X$_2$. In H-stacks, on the contrary, electrons localize at sites C with minimal and maximal contribution of the electron-binding intralayer exciton to X$_1$ and X$_2$. At low doping levels, spatial co-localization of moir\'e-trapped electrons with the intralayer exciton part of moir\'e excitons X$_1$ and X$_2$ thus determines the presence or absence of the corresponding trion states, with binding energies bound by the trion binding energy of $\sim22$~meV in monolayer MoTe$_2$ \cite{Ruppert2014,Lezama2015,Froehlicher2016,Robert2016,Biswas2023} due to finite admixing of interlayer character. 

In summary, our combined experimental and theoretical study of MoTe$_2$-MoSe$_2$ heterostructures identifies the hybrid nature of moir\'e excitons as a function of twist angle. The optically active moir\'e states with resonances in absorption and emission are composed of admixtures of intralayer and interlayer excitons, with strong dependence of the respective fractional contributions on the twist angle. These fractions in turn determine the $g$-factors of moir\'e excitons, which we quantified in magneto-luminescence and in \textit{ab initio} calculations. Moreover, these fractions condition distinct responses of H- and R-type heterostacks to electron doping at low doping levels, where electrons are spatially localized in stacking-specific points of the moir\'e supercell to form bound complexes with co-localized moir\'e excitons. The high degree of intralayer-interlayer hybridization of moir\'e excitons in MoTe$_2$-MoSe$_2$ heterobilayers with near-resonant conduction band alignment, along with their optical signatures of charge carrier doping, establish the system as a versatile material platform for future studies of moir\'e correlated quantum phenomena.

\vspace{11pt}
\noindent \textbf{Associated Content}\\
\noindent \textbf{Supporting Information}: Details on sample fabrication, optical spectroscopy methods, optical spectroscopy of monolayers, determination of heterobilayer twist angles, sample homogeneity, photoluminescence decay of hybrid excitons, excitation power dependence of hybrid excitons, theoretical model of exciton hybridization, dependence of hybrid excitons on twist angle and density functional theory.

\noindent \textbf{Author Information}

\noindent \textbf{Corresponding authors:}\\ X.\,H. (xinhuang@iphy.ac.cn), A.\,H. (alexander.hoegele@lmu.de) and A.\,S.\,B. (anvar.baimuratov@lmu.de). 

\noindent \textbf{Author contributions:}\\
S.\,Z. and X.\,H. contributed equally to this work.

\noindent \textbf{Notes:}\\
The authors declare no competing financial interest. 

\vspace{11pt}
\noindent \textbf{Acknowledgements:} This research was funded by the European Research Council (ERC) under the Grant Agreement No.~772195 as well as the Deutsche Forschungsgemeinschaft (DFG, German Research Foundation) within the Priority Programme SPP~2244 2DMP and the Germany's Excellence Strategy EXC-2111-390814868 (MCQST). S.\,Z. acknowledges support from the Alexander von Humboldt Foundation. X.\,H. acknowledges the National Natural Science Foundation of China (grant No. 62204259). R.\,G. and J.\,M. acknowledge the scientific support and HPC resources provided by the Erlangen National High Performance Computing Center (NHR@FAU) of the Friedrich-Alexander-Universit\"at Erlangen-N\"urnberg (FAU); the hardware was funded by the German Research Foundation (DFG), grant No. 440719683. Z.\,L. was supported by the China Scholarship Council (grant No.~201808140196). S.\,L. and J.\,H. were supported by the Columbia University Materials Science and Engineering Research Center (MRSEC), through NSF grants DMR-1420634 and DMR-2011738. K.\,W. and T.\,T. acknowledge support from the JSPS KAKENHI (Grant Numbers 20H00354 and 23H02052) and World Premier International Research Center Initiative (WPI), MEXT, Japan. A.\,S.\,B. acknowledges funding by the European Union's Framework Programme for Research and Innovation Horizon 2020 under the Marie Sk{\l}odowska-Curie grant agreement No.~754388 (LMUResearchFellows) and from LMUexcellent, funded by the Federal Ministry of Education and Research (BMBF) and the Free State of Bavaria under the Excellence Strategy of the German Federal Government and the L{\"a}nder. A.\,H. acknowledges funding by the Bavarian Hightech Agenda within the EQAP project.

%

\end{document}


\title{Supplemental Information for \\ Hybrid moir\'e excitons and trions in twisted MoTe$_2$-MoSe$_2$ heterobilayers}
    
    \author{Shen Zhao$^{1,\ddag}$, Xin Huang$^{2,3,1,*,\ddag}$, Roland Gillen$^{4}$, Zhijie Li$^{1}$, Song Liu$^{5}$, Kenji Watanabe$^{6}$, Takashi Taniguchi$^{7}$, Janina Maultzsch$^{4}$, James Hone$^{5}$, Alexander H\"ogele$^{1,8,*}$ and Anvar~S.~Baimuratov$^{1,*}$\footnote[0]{$^{\ddag}$ S.~Z. and X.~H. contributed equally to this work.}\footnote[0]{$^{*}$X. H. (xinhuang@iphy.ac.cn), A. H. (alexander.hoegele@lmu.de), A. S. B. (anvar.baimuratov@lmu.de).}}    
    
    \affiliation{$^1$Fakult\"at f\"ur Physik, Munich Quantum Center, and Center for NanoScience (CeNS), Ludwig-Maximilians-Universit\"at M\"unchen, Geschwister-Scholl-Platz 1, 80539 M\"unchen, Germany}

    \affiliation{$^2$Beijing National Laboratory for Condensed Matter Physics, Institute of Physics, Chinese Academy of Sciences and School of Physical Sciences, CAS Key Laboratory of Vacuum Physics, University of Chinese Academy of Sciences, Beijing 100190, P. R. China}
    
    \affiliation{$^3$School of Physical Sciences, CAS Key Laboratory of Vacuum Physics, University of Chinese Academy of Sciences, Beijing 100190, P. R. China}    
    
    \affiliation{$^4$Department of Physics, Friedrich-Alexander Universit\"at Erlangen-N\"urnberg, Staudtstra{\ss}e 7, 91058 Erlangen, Germany}
    
    \affiliation{$^5$Department of Mechanical Engineering, Columbia University, New York, New York 10027, United States}
    
    \affiliation{$^6$Research Center for Electronic and Optical Materials, National Institute for Materials Science, 1-1 Namiki, Tsukuba 305-0044, Japan}
    
    \affiliation{$^7$Research Center for Materials Nanoarchitectonics, National Institute for Materials Science, 1-1 Namiki, Tsukuba 305-0044, Japan}
    
    \affiliation{$^8$Munich Center for Quantum Science and Technology (MCQST), Schellingstra\ss{}e 4, 80799 M\"unchen, Germany}

\maketitle

\vspace{-10pt}

\vspace{20pt}

\tableofcontents
\clearpage

\noindent \textbf{Supplementary Note $\mathbf{1}$: Sample fabrication}
\phantomsection
\addcontentsline{toc}{section}{Note $\mathbf{1}$: Sample fabrication}
\vspace{8pt}

Fully hexagonal boron nitride (hBN) encapsulated MoTe$_2$-MoSe$_2$ heterostructure devices were fabricated by a dry-transfer method~\cite{Pizzocchero2016}. First, monolayers of MoTe$_2$, MoSe$_2$, few layer graphite and thin flakes of hBN were mechanically exfoliated from their bulk crystals onto Si/SiO$_2$ substrates. MoTe$_2$ bulk crystals were synthesized using a two-step flux method~\cite{Liu2023} with a relatively low defect density ($<10^{10}$~cm$^{-2}$). MoSe$_2$ and graphite bulk crystals were purchased from HQ graphene, while hBN crystals were provided by the National Institute for Materials Science (NIMS). In a second step, a PC/PDMS stamp was used to assemble atomically thin flakes layer-by-layer. The pick-up temperatures for hBN flakes, thin graphite layers and monolayers of MoTe$_2$ and MoSe$_2$ was 50~$^\circ$C, 90~$^\circ$C, 40~$^\circ$C, and 50~$^\circ$C, respectively. The entire heterostructure was subsequently released from the stamp onto a 285~nm Si/SiO$_2$ substrates with prepatterned Cr/Au (5~nm/40~nm) electrodes at 170~$^\circ$C. Finally, the device was annealed in high vacuum ($10^{-9}$~mbar) for 12~h at a temperature of 200~$^\circ$C to remove polymer residuals and improve interlayer coupling~\cite{Alexeev2017}. After the fabrication, the stacking type and twist angle of each MoTe$_2$-MoSe$_2$ heterobilayer was determined by second harmonic generation measurements (see Supplementary Note~4).

\vspace{12pt}
\noindent \textbf{Supplementary Note $\mathbf{2}$: Optical spectroscopy methods}
\phantomsection
\addcontentsline{toc}{section}{Note $\mathbf{2}$: Optical measurement methods}
\vspace{8pt}

All optical measurements were performed at 3.2~K using a home-built confocal microscope in back-scattering geometry. The samples were loaded into a closed-cycle cryostat (attocube systems, attoDRY1000) with magnetic field ranging from $-9$~T to $+9$~T in the Faraday configuration; $xyz$ piezo-stepping and $xy$ scanning units (attocube systems, ANPxyz and ANSxy100) were employed for sample positioning. An apochromatic cryogenic objective with a numerical aperture of 0.81 (attocube systems, LT-APO/IR/0.81) was used to confocally excite and collect the signal from sample with a spot size of $\sim$1~\micro\meter. For photoluminescence (PL) measurements, the excitation source was a Ti:sapphire laser (Coherent, Mira) in continuous-wave mode. The laser was tuned to the $2s$ exciton resonance of monolayer MoTe$_2$ at 975~nm and the incident power was kept at 5~\micro\watt~unless otherwise specified. For differential reflectivity (DR) measurements, a power-stabilized Tungsten-Halogen lamp (Thorlabs, SLS201L) was used as a broadband light source. The PL or reflection signal were spectrally dispersed by a monochromator (Roper Scientific, Acton SP2500 with a 300 grooves/mm grating) and detected by a liquid nitrogen cooled charge-coupled device (Roper Scientific, Spec-10:100BR) or a one-dimensional linear InGaAs photodiode array (Roper Scientific, OMA V:1024-1.7 LN). The DR spectra were obtained by normalizing the reflected spectrum of the sample ($R_{\textrm{s}}$) to the reference spectrum from nearby region without MoTe$_2$ and MoSe$_2$ layers but having both hBN and graphite layers ($R_{\textrm{r}}$), as $\textrm{DR} =  \Delta R/R = (R_{\textrm{s}}-R_{\textrm{r})}/R_{\textrm{r}}$. The absorption $\chi''$ spectra were then derived from DR by accounting for the interference of multiple thin-layer reflections \cite{Back2017,Zhao2023}. For magneto-PL measurements, a set of linear polarizers (Thorlabs, LPVIS), half- and quarter-waveplates (B. Halle, super-achromatic) mounted on piezo-rotators (attocube systems, ANR240) were used to control the polarization in excitation and detection. For time-resolved PL measurements, a supercontinuum laser (NKT Photonics, SuperK Extreme) with a pulse duration of 4~ps was coupled to an acousto-optic tunable filter (NKT Photonics, SuperK Select nIR1) for pulsed excitation at 975~nm. The PL signal was detected by a superconducting nanowire single-photon detector (Scontel, TCOPRS-CCR-SW-85) and the photon detection events were time-correlated with the trigger signal from the supercontinuum laser by a electronic correlator (PicoQuant, PicoHarp 300). The count rate was adjusted to $<1\%$ of the laser rate to prevent pile-up effects. The instrument response function (IRF) was measured by detecting the laser pulse using the same setup, and the bi-exponential fitting of the PL decay with deconvoluted IRF were performed using EasyTau 2 software (Picoquant).  To control charge carrier density, the voltage was applied to the graphite gate with a sourcemeter (Keithley 2400). 

\vspace{12pt}
\noindent \textbf{Supplementary Note $\mathbf{3}$: Optical spectroscopy of monolayers}
\phantomsection
\addcontentsline{toc}{section}{Note $\mathbf{3}$: Optical spectroscopy of monolayers}
\vspace{8pt}

\begin{figure*}[t!]
    \includegraphics[scale=1]{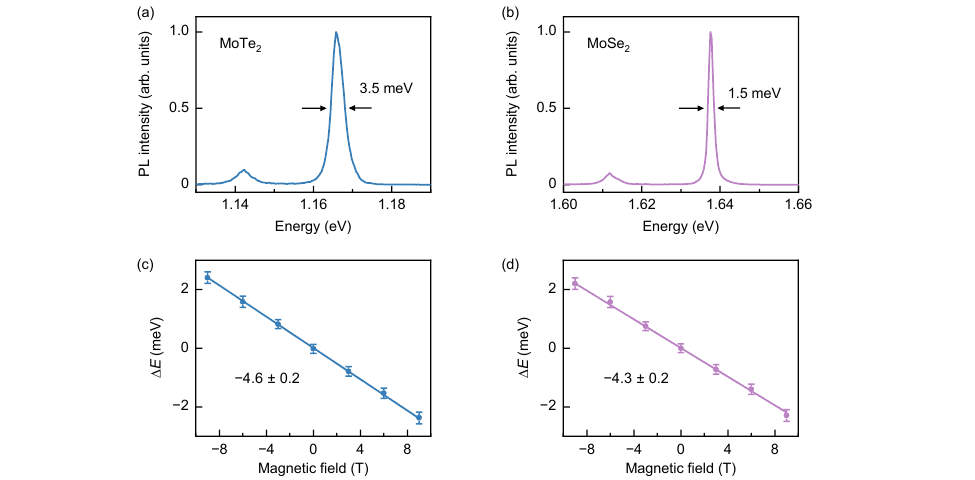}
    \caption{PL characterization of MoTe$_2$ and MoSe$_2$ monolayers in a MoTe$_2$-MoSe$_2$ heterobilayer sample. (a) and (b) Typical PL spectra of MoTe$_2$ and MoSe$_2$ at zero gate voltage, showing narrow linewidth of the A-exciton transitions and weak emission from trions. (c) and (d) Zeeman splitting $\Delta E$ of MoTe$_2$ and MoSe$_2$ A-excitons as a function of the magnetic field. The corresponding $g$-factors are extracted from the slopes of linear fits shown by solid lines.}
    \label{monolayer}
\end{figure*}

Figure~\ref{monolayer} displays the PL characteristics results of monolayer regions in an hBN-encapsulated device. For both MoTe$_2$ and MoSe$_2$, their typical PL spectra show bright and spectrally narrow A-exciton peaks. The corresponding full width at half maximum (FWHM) linewidths are 3.5 and 1.5~meV, respectively, approaching the homogeneous limit~\cite{Cadiz2017}. Moreover, the spectra have very weak trion features as compared to the A-exciton, which implies a low density of charge defects~\cite{Edelberg2019}. These results confirm high quality of the individual monolayers upon device fabrication. To compare the exciton $g$-factors between monolayer and heterobilayer, we performed magneto-photoluminescence measurements on the A-exciton of monolayer MoTe$_2$ and MoSe$_2$. The obtained $g$-factors are respectively $-4.6$ and $-4.3$, in accord with previous reports~\cite{Arora2016,Goryca2019}.

\vspace{14pt}
\noindent \textbf{Supplementary Note $\mathbf{4}$: Determination of heterobilayer twist angles}
\phantomsection
\addcontentsline{toc}{section}{Note $\mathbf{4}$: Determination of heterobilayer twist angles}
\vspace{8pt}

\begin{figure*}[t!]
    \includegraphics[scale=1]{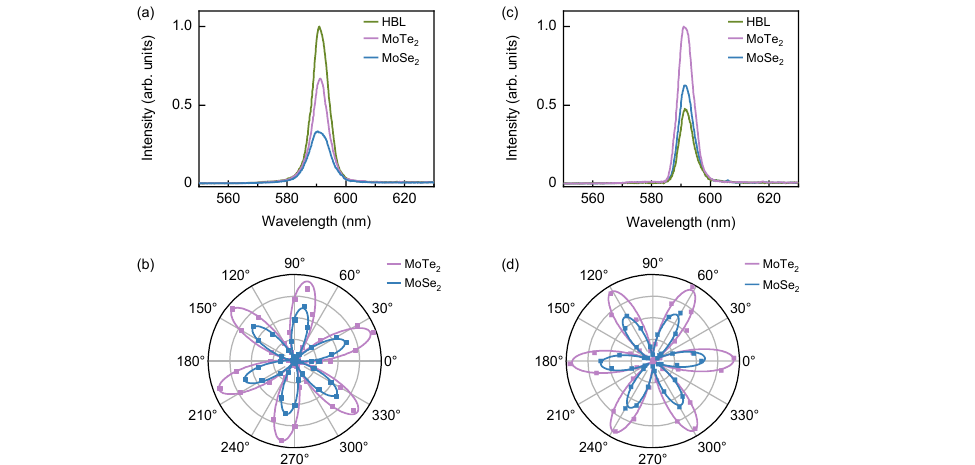}
    \caption{Twist-angle measurements using second-harmonic generation. (a) SHG signal of an R-type heterobilayer measured on the monolayer MoSe$_2$, monolayer MoTe$_2$ and heterobilayer (HBL) parts with the same experimental configuration. (b) Polarization-resolved SHG intensity of monolayer MoSe$_2$ and MoTe$_2$ parts. The solid lines are the fits with sinusoidal functions. (c) and (d) Same measurements for an H-type heterobilayer sample.}
    \label{SHG}
\end{figure*}

To determine the relative twist angle between MoTe$_2$ and MoSe$_2$ layers, we performed polarization-resolved second harmonic generation (SHG) measurements at 3.2~K using a home-built system. A supercontinuum laser with 4~ps pulse duration and 80~MHz repetition rate was used to excite at 1280~nm below the optical bandgap of MoTe$_2$-MoSe$_2$ heterobilayer. For each measurement, the excitation laser with a power of 5~mW was linearly polarized and focused onto a $\sim2$~\micro\meter~spot by an objective lens with a numerical aperture of 0.81. The SHG signal was collected by the same objective and separated from the reflected laser using a beam splitter and a short-pass filter. A linear polarizer was then placed in the SHG signal path and aligned parallel to the excitation polarization as analyzer. The angle resolution was realized by rotating a motorized half-wave plate located just above the objective. The SHG signal was finally recorded by a spectrometer with a 300 grooves/mm grating and a liquid-nitrogen-cooled charge-coupled device.

Given the six-fold rotational symmetry of monolayer transition metal dichalcogenides, the case of R-type ($\sim0^\circ$) and H-type ($\sim60^\circ$) bilayers cannot be differentiated by separately measuring the angle-resolved SHG of each constituent monolayer. To identify the type of  R- or H-stacking, one has to additionally measure the SHG from the bilayer region. As exemplified in Fig.~\ref{SHG} (a) and (c), for the R-type (H-type) case the second harmonic fields of MoTe$_2$ and MoSe$_2$ layers constructively (destructively) interfere, making the SHG signal of the heterobilayer region stronger (weaker) than in the individual monolayer regions for the same experimental configuration. Note that the SHG signal of monolayer MoTe$_2$ is much stronger than that of monolayer MoSe$_2$, consistent with the higher second-order nonlinearity in compounds of telluride than that in compounds of selenide~\cite{Shen1984}. With the information on the stacking type, the angle-resolved SHG intensity patterns were fitted with $I_{\text{SHG}} \propto \text{cos}^2[3(2\alpha - \phi)]$, where $\alpha$ is the rotation angle of the half-wave plate and $\phi$ is a fitting parameter defining the armchair orientation of the crystal. Thus, for R- and H-type heterobilayer, the twist angle $\theta$ between the two layers is determined by $\theta=\mid \phi_1 - \phi_2 \mid$ and $\theta= 60^\circ - \mid \phi_1 - \phi_2 \mid$, with $\phi_1$ and $\phi_2$ obtained from the fitting of the respective monolayer regions. For the two representative heterobilayers in Fig.~\ref{SHG}, the twist angles were determined as $2.0^\circ$ and $58.1^\circ$, respectively. The twist angle uncertainty of this method was estimated to be $\sim0.2^\circ$, from fitting uncertainties of the SHG patterns.


\vspace{16pt}
\noindent \textbf{Supplementary Note $\mathbf{5}$: Sample homogeneity}
\phantomsection
\addcontentsline{toc}{section}{Note $\mathbf{5}$: Sample homogeneity}
\vspace{8pt}

\begin{figure*}[b!]
    \includegraphics[scale=1]{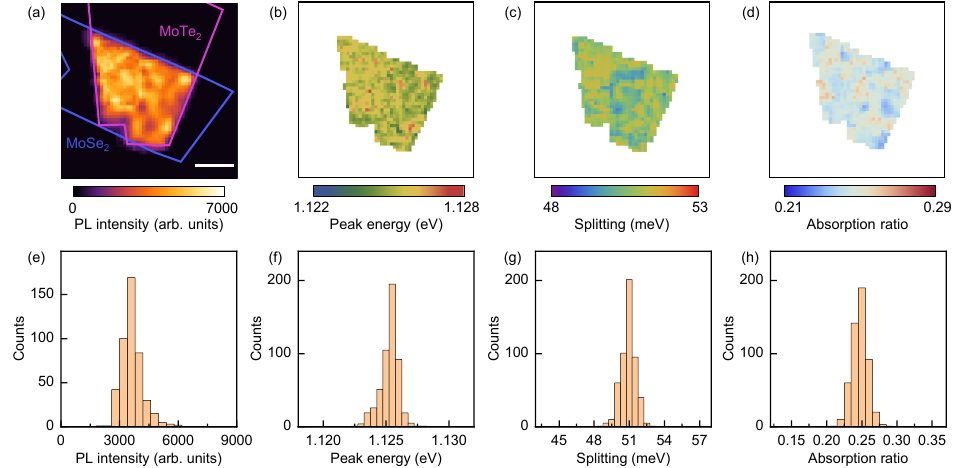}
    \caption{Spatial maps of PL intensity (a) and peak energy (b) for X$_1$ exciton as well as the energy splitting (c) and absorption strength ratio (d) between X$_1$ and X$_2$ excitons, measured on the heterobilayer with $2^\circ$ twist angle. The histograms in panels (e--h) show the statistics extracted from the maps in (a--d), respectively. The average PL intensity of peak X$_1$ in arbitrary units is 3843, with a standard deviation of 585. The average peak energy of X$_1$ and energy  splitting between X$_1$ and X$_2$ are 1.125~eV and 51~meV, respectively, with the corresponding standard deviations of 2.1 and 1.8~meV. The average absorption strength ratio is 0.25 with a standard deviation of 0.02. The positions at the heterobilayer edges were excluded from the statistics. Scale bar in (a), 3~\micro\meter.}
    \label{homogeneity_R}
\end{figure*}

\begin{figure*}[t!]
    \includegraphics[scale=1]{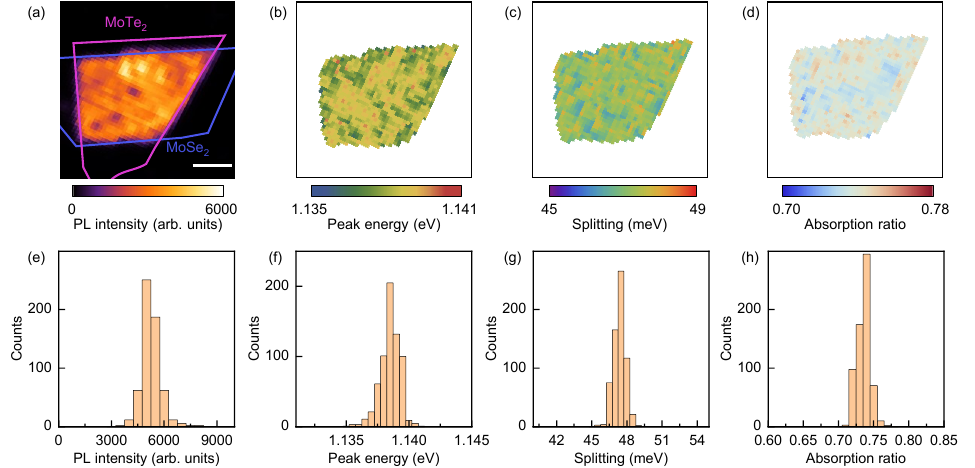}
    \caption{Same as Fig.~\ref{homogeneity_R}, measured on the heterobilayer with $58^\circ$ twist angle. The average PL intensity (in arbitrary units), peak energy, energy splitting and absorption strength ratio are 5348, 1.138~eV, 47~meV and 0.74, respectively, with the corresponding standard deviations of 523, 2.7~meV, 1.9~meV and 0.02. Scale bar in (a), 4~\micro\meter.}
    \label{homogeneity_H}
\end{figure*}

For all samples studied in this work, we performed hyperspectral mapping of spatial homogeneity. Figure~\ref{homogeneity_R} and \ref{homogeneity_H} show the results of a representative aligned R-type and H-type samples. Throughout the entire heterobilayer regions, the PL and absorption features vary only marginally with characteristics of X$_1$ and X$_2$ excitons. From the statistical analysis of the maps we find that the PL intensities of X$_1$ are relatively uniform with a typical fluctuation below $30\%$, and for both X$_1$ and X$_2$ peaks the dispersions of the energy positions are smaller than 3~meV. Moreover, the variations of the energy splitting between the two peaks are also small ($<2$~meV), and the same holds for their absorption strength ratio ($<15\%$). The above results confirm the rather uniform optical characteristics of our devices, and rule out the formation of reconstructed domains in MoTe$_2$-MoSe$_2$ heterobilayers signified by drastic spatial variations in the sample~\cite{Shabani2021,Zhao2023}. The experimental data in Fig.~1(e) and (f) of the main text were obtained from statistical averaging of the corresponding spatial maps, and their standard deviations are represented as error bars. 


\vspace{12pt}
\noindent \textbf{Supplementary Note $\mathbf{6}$: Photoluminescence decay of hybrid excitons}
\phantomsection
\addcontentsline{toc}{section}{Note $\mathbf{6}$: Photoluminescence decay of hybrid excitons}
\vspace{8pt}

\begin{figure*}[t!]
    \includegraphics[scale=1]{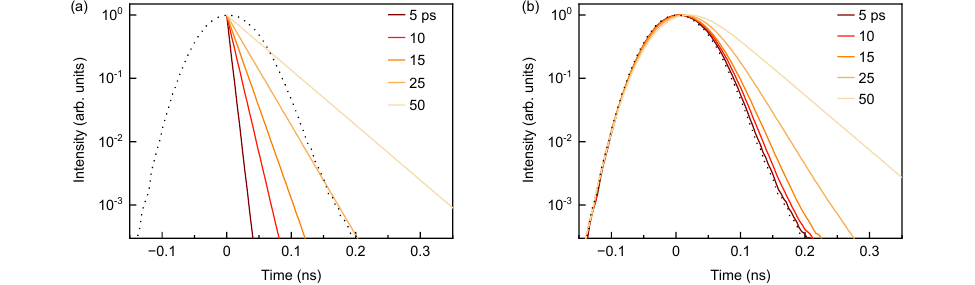}
    \caption{Estimation of the temporal resolution for lifetime measurements. (a) and (b) Mono-exponential decay traces for different time constants (solid lines) before and after convolution with the instrument response function (dotted line), respectively. For time constant shorter than 10~ps, the decay traces become indistinguishable from the instrument response function.}
    \label{lifetime_resolution}
\end{figure*}

To study the recombination dynamics of hybrid excitons, we performed time-resolved PL measurements on the lowest X$_1$ exciton transition for various heterobilayer twist angles. As the hybrid excitons have sizeable oscillator strength, their lifetimes become very short and comparable with the instrument response function (IRF) of our setup. To analyze the decay, we carried out iterative reconvolution fit by taking into account the IRF. In this way, we were able to determine measure lifetimes significantly smaller than the IRF~\cite{OConnor1984}. As illustrated in Fig.~\ref{lifetime_resolution}, the estimated temporal resolution limit of our setup using this fitting method is $\sim$10~ps (compared to 100~ps for the FWHM of IRF). For shorter lifetimes, the decay traces 
overlap increasingly with the IRF, rendering the fit less reliable. 

\begin{figure*}[b!]
    \includegraphics[scale=1]{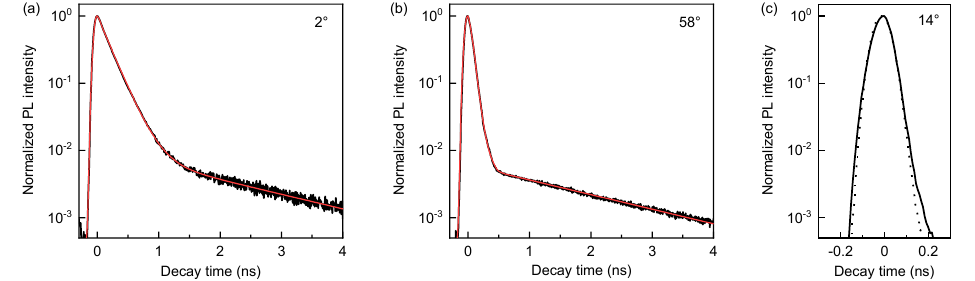}
    \caption{Full-time-scale photoluminescence decays of hybrid X$_1$ exciton of heterobilayers with twist angles of $2^\circ$ (a), $58^\circ$ (b) and $14^\circ$ (c). The red lines in (a) and (b) represent bi-exponential decay convoluted with the IRF as best fits. The dotted line in (c) corresponds to the IRF.}
    \label{lifetime_long}
\end{figure*}

For almost aligned heterobilayers ($0^\circ\le\theta\le7^\circ$ and $53^\circ\le\theta\le60^\circ$), the lifetimes are long enough and the decay traces can be distinguished from the IRF. We used the reconvolution fitting method with bi-exponential function to fit these PL decays. As exemplified in Fig.~\ref{lifetime_long}(a) and (b), our fitting captures both the short and the long decay components with fitting residuals always below $5\%$ and we found that the short component are always dominant (weight $>95\%$) with a small contribution from the long component. For misaligned heterobilayers ($7^\circ<\theta<53^\circ$), the decay traces is very short and overlaps with the IRF, as seen in Fig.~\ref{lifetime_long}(c). To a first approximation, we consider them as mono-exponential decays with the decay time equal to the resolution limit of 10~ps (on the same order of radiative lifetime for monolayer A-exciton~\cite{Fang2019}).

\begin{figure*}[t!]
    \includegraphics[scale=1]{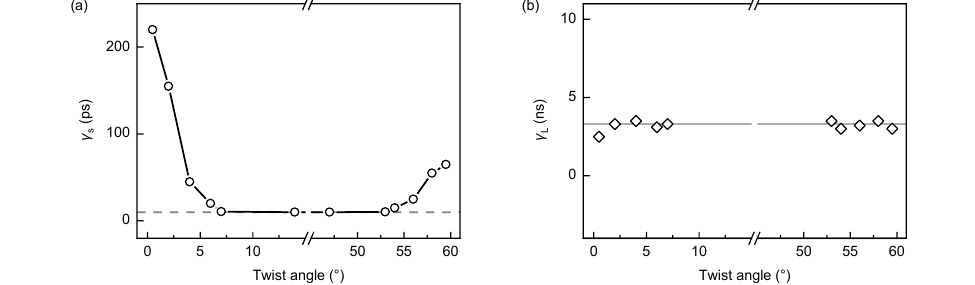}
    \caption{(a) and (b) Short and long decay times of the X$_1$ exciton as a function of twist angle, respectively. Since the decay traces in misaligned heterobilayers ($7^\circ<\theta<53^\circ$) are very short and overlap significantly with the IRF. We plotted the upper limit of their decay time in (a), which corresponds to the estimated temporal resolution limit of the setup (indicated by the dashed line).}
    \label{lifetime_summary}
\end{figure*}

Figure~\ref{lifetime_summary} summarizes the obtained characteristic decay times $\gamma_\text{S}$ and $\gamma_\text{L}$(corresponding to the short and long decay component, respectively) for different twist angles. By comparing with the plot in Fig.~1(f) of the main text, we found that the short decay time $\gamma_\text{S}$ scales inversely with the oscillator strength of X$_1$, and we thus associate this dominant decay component with the radiative recombination channel. In contrast, the long decay time $\gamma_\text{L}$ remains relatively similar for all heterobilayer twist angles. We anticipated that this long component (several nanoseconds) with marginally contribution correspond to the non-radiative processes to the momentum-dark reservoir at the $\kappa$ point~\cite{Miller2017}.

\clearpage

\clearpage


\noindent \textbf{Supplementary Note $\mathbf{7}$: Excitation power dependence of hybrid excitons}
\phantomsection
\addcontentsline{toc}{section}{Note $\mathbf{7}$: Excitation power dependence of hybrid excitons}
\vspace{8pt}

\begin{figure*}[h!]
    \includegraphics[scale=1]{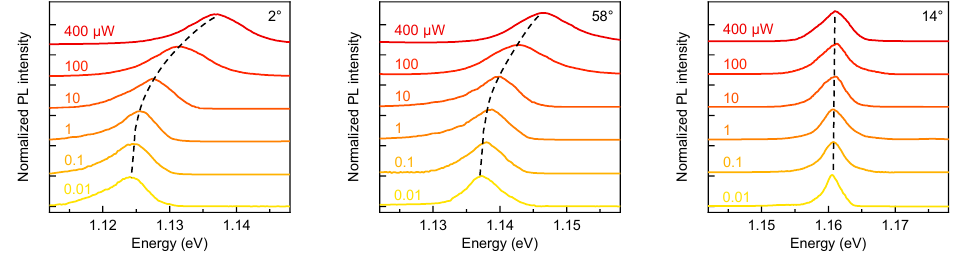}
    \vspace{-1cm}
    \caption{Power-dependent PL spectra for the X$_1$ exciton in $2^\circ$, $58^\circ$ and $14^\circ$ twisted heterobilayers. The dashed lines highlight the energy evolution of peak maximum with power increasing. For the hybrid cases ($\theta = 2^\circ$ and $58^\circ$), the X$_1$ peak shows a continuous energy blue-shift with increased power by up to 15 and 10~meV, receptively. By comparison, for the non-hybrid case ($\theta = 14^\circ$), the peak remains almost non-shifted.}
    \label{power_PL}
\end{figure*}

\begin{figure*}[h!]
    \includegraphics[scale=1]{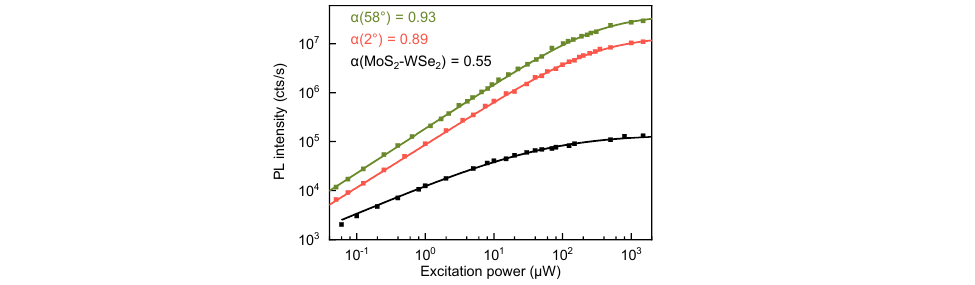}
    \vspace{-1cm}
    \caption{Comparison of power-dependent PL intensity between the hybrid X$_1$ excitons in aligned MoTe$_2$-MoSe$_2$ heterobilayers ($\theta = 2^\circ$ and $58^\circ$, red and green) and the pure interlayer exciton in an aligned MoS$_2$-WSe$_2$ heterobilayer (black). The solid curves are fits to data (filled squares) using the saturation function $I=I_{\text{sat}}/(1+P_{\text{sat}}/P^{\alpha})$. We obtained almost linear scaling $\alpha = 0.89$ and 0.93 with the saturation powers $P_{\text{sat}}$ of 160 and 210~\micro\watt~and the saturation intensities $I_{\text{sat}}$ of $1.3\times10^7$ and $3.9\times10^7$ counts~s$^{-1}$ for the hybrid excitons in $2^\circ$ and $58^\circ$ MoTe$_2$-MoSe$_2$, respectively. In contrast, for the interlayer exciton in MoS$_2$-WSe$_2$, the intensity starts to show a sub-linear response ($\alpha=0.55$) from a very low power of 0.5~\micro\watt~and becomes saturated around 10~\micro\watt~with a saturation intensity of $2.8\times10^5$ counts~s$^{-1}$, which is much lower than the hybridized excitons by a factor of $\sim$100. The measurements were conducted using the same setup.}
    \label{power_intensity}
\end{figure*}

\clearpage

\noindent \textbf{Supplementary Note $\mathbf{8}$: Theoretical model of exciton hybridization}
\phantomsection
\addcontentsline{toc}{section}{Note $\mathbf{8}$: Theoretical model of exciton hybridization}
\vspace{8pt}

\begin{figure*}[b!]
    \includegraphics[scale=1.5]{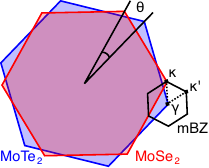}
    \caption{Formation of a mini Brillouin zone in MoTe$_2$-MoSe$_2$ heterobilayer.
}
    \label{mbz}
\end{figure*}

To model the experimental results, we applied the continuum model for near-resonant interlayer hybridization ~\cite{Polovnikov2024,RuizFalko2019,MacdonaldTopo2017}. The MoTe$_2$-MoSe$_2$ heterobilayers facilitate the analysis of the lowest-energy excitons due to their substantial valence band offset $\sim 0.5$~eV~\cite{Zhang2016}. On the other hand, their conduction bands are nearly aligned, therefore we focus on MoTe$_2$ intralayer excitons (X) and interlayer excitons (IX), which consist of a hole in MoTe$_2$ and an electron in MoSe$_2$. The twist angle $\theta$ and the lattice constant mismatch between two layers induce a valley mismatch represented by vectors 
$\Delta\mathbf{K} = \mathbf{K}_\text{Se} - \mathbf{K}_\text{Te}$ and $\Delta\mathbf{K} = \mathbf{K}'_\text{Se} - \mathbf{K}_\text{Te}$ for R- and H-type heterobilayers, respectively. The emergent periodicity of the moir\'e superlattice leads to the formation of a mini Brillouin zone (mBZ) with reciprocal basis vectors
\begin{equation}
    \mathbf{b}_{1,2} = \frac{\sqrt 3}{2}
    \begin{bmatrix}
      \sqrt 3 & \pm 1 \\
      \mp 1 & \sqrt 3
    \end{bmatrix}
    \Delta\mathbf{K}.
\end{equation}
We define the mBZ as shown in Fig.~\ref{mbz} with the center $\gamma$ at the $K$-valley of MoTe$_2$ ($\mathbf{K}_\text{Te}$), and the point $\kappa$ at the $K$- or $K'$-valley of MoSe$_2$. With this definition of the mBZ, the momentum-direct excitons are located at the $\gamma$ point. 

Following the model from Ref.~\citenum{Polovnikov2024}, we introduce the Hamiltonian 
\begin{equation}
    H = \\
    \begin{pmatrix}
      H_\text{X} & T \\
      T^*& H_\text{IX}
    \end{pmatrix},
    \label{ham}
\end{equation}
where we assume parabolic dispersions of intralayer and interlayer excitons, with the corresponding diagonal blocks are written as follows:
\begin{align}
    \langle \text{X}, \mathbf{k} + \mathbf{g'}| H_\text{X}| \text{X}, \mathbf{k + g} \rangle =&
        \delta_{\mathbf{g,g'}} \left( E_\text{X} + \frac{\hbar^2 |\mathbf{k + g}|^2}{2M_\text{X}} \right)
        + \sum_{j = 1}^6 V_j \delta_{\mathbf{g-g',G}_j} ,
    \\
    \langle \text{IX}, \mathbf{k'} + \mathbf{g'}| H_\text{IX} | \text{IX}, \mathbf{k' + g} \rangle =&
    \delta_{\mathbf{g,g'}} \left( E_\text{IX} + \frac{\hbar^2 |\mathbf{k + g}'|^2}{2M_\text{IX}} \right),
    \label{IXen}
\end{align}
where the reciprocal vectors are $\mathbf{g} = n \mathbf{b}_1 + m \mathbf{b}_2$, $n$ and $m$ are integers, $\mathbf{k}$ and $\mathbf{k}'$ are the center-of-mass wave vectors of X and IX excitons measured from $\gamma$ and $\kappa$, respectively; $E_\text{X(IX)}$ and $M_\text{X(IX)}$ are the respective energies and effective masses; $V_1 = V_3 = V_5 \equiv V$ and $V_2 = V_4 = V_6 \equiv V^*$ characterize the lowest-order harmonic expansion of the electron moir\'e potential in MoTe$_2$ with the first-shell reciprocal lattice vectors  $\mathbf G_j$ ($j = 1,2,...,6$). We assume that the moir\'e potentials for holes in MoTe$_2$ and for electrons in MoSe$_2$ are relatively small and we neglect their contributions in model~\cite{Polovnikov2024}.
In Table~\ref{massgap} we summarize the relevant parameters for the singlet and triplet excitons, which are further considered independently.

\begin{table}[!b]
    \begin{ruledtabular}    
    \begin{tabular}{rrrrrrrr}  
        type & spin state & $E_\text{X}$ & $E_\text{IX}$ & $M_\text{X}$ & $M_\text{IX}$ \\  
        \hline
        R & singlet       & $E_\text{0}$ & $E_0 + \Delta$ & $m_c + m_v$ & $m_c' + m_v$ \\
        R & triplet       & $E_0 + \Delta_\mathrm{SO}$ & $E_0 + \Delta + \Delta_\mathrm{SO}'$ &  $m_{c+1} + m_v$ &  $m_{c+1} ' + m_v$ \\
        \hline
        H & singlet       & $E_0$ & $E_0 + \Delta + \Delta_\mathrm{SO}'$  & $m_c + m_v$ & $m_{c+1} ' + m_v$ \\
        H & triplet       & $E_0 + \Delta_\mathrm{SO}$ & $E_0 + \Delta$ & $m_{c+1} + m_v$ & $m_c' + m_v$  \\
    \end{tabular}
    \caption{Energies and effective masses of intralayer and interlayer excitons. $E_0$ is the A-exciton energy in MoTe$_2$ monolayer, and $\Delta$ is the energy offset between interlayer and intralayer excitons. The effective mass parameters $m_c$ , $m_{c+1}$, $m_v$ and the spin-orbit splitting $\Delta_\mathrm{SO}$ correspond to the MoTe$_2$ layer, whereas the effective mass parameters $m_c'$, $m_{c+1}'$ and the spin-orbit splitting $\Delta_\mathrm{SO}'$ are related to the MoSe$_2$ layer. In our model, $E_0$ and $\Delta$ are fitting parameters, and all other parameters are obtained from our \textit{ab initio} calculations (see Table~\ref{parameters}).}
    \label{massgap}
    \end{ruledtabular}
\end{table}

The tunneling of conduction band electrons is described by interlayer hopping elements
\begin{equation}
    \langle \text{IX}, \mathbf{k' + g'}|
    T|
    \text{X}, \mathbf{k + g} \rangle =
    t (\delta_{\mathbf{k+g},\mathbf{k'+g'} + \Delta \mathbf{K}}+ 
       \delta_{\mathbf{k+g},\mathbf{k'+g'} + C^1_3 \Delta \mathbf{K}}+
       \delta_{\mathbf{k+g},\mathbf{k'+g'} + C^2_3 \Delta \mathbf{K}}),
\end{equation}
with the hopping parameter $t$, and rotation matrix $C_\nu^\mu$ with angle $2\pi \mu/\nu$. 

The diagonalization of the Hamiltonian in Eq.~(\ref{ham}) provides eigenenergies and eigenvectors of hybrid moir\'e states. We assume that their oscillator strengths are proportional to the squared norm of their projections onto the fundamental A-exciton state. In our analysis, $E_0$, $\Delta$, $V$ and $t$ are treated as free parameters, and other band parameters are obtained from our \textit{ab initio} calculations (see Supplementary Note~10). By fitting the model to twist-dependent data shown in Figs.~1(c--f), we reproduce the energies and oscillator strengths of the two moir\'e excitons with very good agreement, with $E_0 = 1162$~meV, $\Delta = -54$~meV, and $t = 12$~meV. 

For the fitting procedure, we employed the method of least squares and simultaneously minimized the respective squares of the exciton energies $E_i$ and oscillator strengths $f_i$. With dimensionless energies, obtained from normalization by the spectral window of $40$~meV, we minimized the function $\mathcal{F} = \sum_{\theta_j} \left[ [(E-E_i)/40~\text{meV}]^2 + (f-f_i)^2 \right]$, where we sum over all experimental values of twist angles $\theta_j$. Intralayer potentials for R-type and H-type stackings have the same amplitude $|V| = 1$~meV, but different phases, $\arg (V) = 3 \pi /4$ and $7 \pi /12$, respectively. In our fitting procedure, we find that the parameter $|V|$ in the range between $0$ and $3$~meV leads to approximately the same residuals. To obtain the best-fit value for this parameter, we take into account the constrains of phases $\arg (V)$, depending on the trion binding energy results from the charge-doping experiments on $2^\circ$ and $58^\circ$ samples.

\begin{figure*}[b!]
    \includegraphics[scale=1.1]{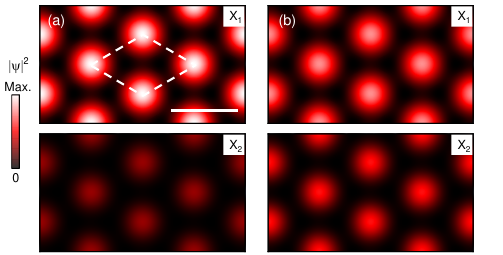}
    \vspace{-0.5cm}
    \caption{Color-coded maps of real-space probability functions $|\psi|^2$ of interlayer excitons contributing to X$_1$ (upper panel) and X$_2$ (lower panel) excitons in heterostacks with twist angles of (a) $2^\circ$ and (b) $58^\circ$. The dashed rhombus indicates the moir\'e untit cell, the scale bars are $5$~nm.}
    \label{IX_wf}
\end{figure*}

In Fig.~4(c) and (d) of the main text, we plot the potential linecuts along the main diagonal of the moir\'e unit cell using the following real-space representation:
\begin{equation}
     \mathcal V (\mathbf r) = \sum_{j = 1}^6 V_j e^{i \mathbf G_j \mathbf r}.
\end{equation}
In the same way we compute the real-space distributions of the intralayer contributions to the hybrid moir\'e excitons~\cite{Polovnikov2024}, shown in Fig.~4(e) and (f) of the main text for X$_1$ $(m = 1)$ and X$_2$ $(m = 2)$ states in $\theta = 2^\circ$ and $58^\circ$ heterobilayers:
\begin{equation}
    \psi_{m}(\mathbf{r})  = \sum_{\mathbf G_l \in \mathbf G} c_{lm} e^{i \mathbf{G}_l \mathbf{r}},
\end{equation}
where we use the obtained eigenvectors $c_{lm}$ and sum over corresponding wave vectors of intralayer states in the calculation of $\mathbf G$.

Accordingly, in Fig.~\ref{IX_wf} we show the real-space distributions of the interlayer contributions to hybrid moir\'e excitons obtained from the following equation:
\begin{equation}
    \psi_{m}^\text{IX} (\mathbf{r})  = \sum_{\mathbf G_l \in \mathbf G^\text{IX}} c_{lm}^\text{IX} e^{i \mathbf{G}_l \mathbf{r}},
\end{equation}
where the eigenvectors $c_{lm}^\text{IX}$ and wave vectors $\mathbf G^\text{IX}$ correspond to the interlayer states.

\vspace{16pt}
\noindent \textbf{Supplementary Note $\mathbf{9}$: Dependence of hybrid excitons on twist angle}
\phantomsection
\addcontentsline{toc}{section}{Note $\mathbf{9}$: Dependence of hybrid excitons on twist angle}
\vspace{8pt}

\begin{figure*}[t!]
    \includegraphics[width=\textwidth]{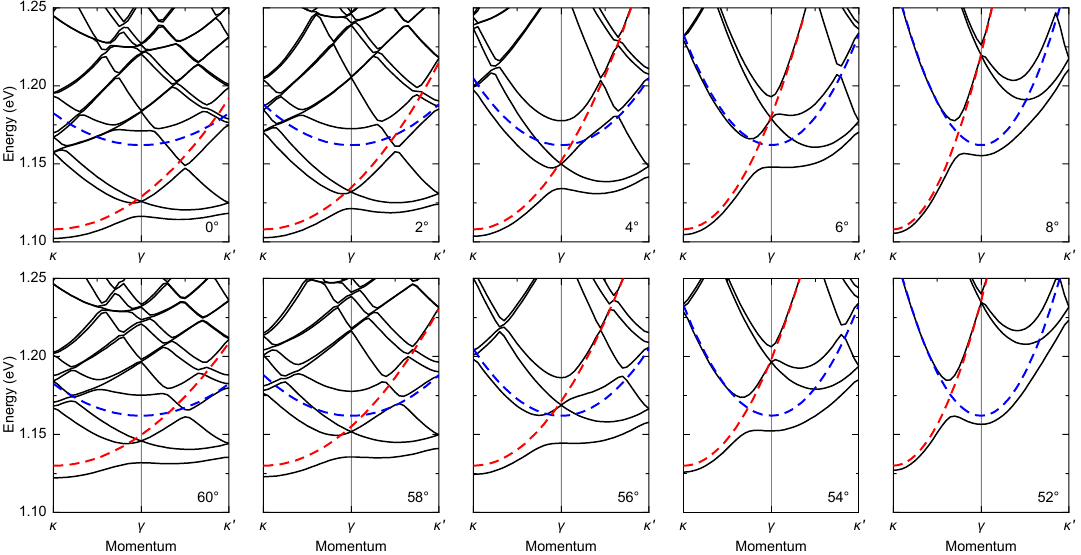}
    \caption{Moir\'e band structures of hybridized excitons in MoTe$_2$-MoSe$_2$ heterobilayers calculated for different interlayer twist angles. The panels depict only spin-bright (singlet) hybridized exciton bands within the first moir\'e Brillouin zone. Red and blue dashed lines in all panels represent the uncoupled MoTe$_2$ A-exciton and the interlayer exciton IX states.}
    \label{mbs}
\end{figure*}

Figure~\ref{mbs} displays the calculated hybrid exciton band structures together with the decoupled MoTe$_2$ A-exciton (X$_\text{A}$) and spin-singlet interlayer exciton (IX) bands for various twist angles. Owing to the small conduction band offset between MoTe$_2$ and MoSe$_2$, the two decoupled bands cross inside the mini Brillouin zone. In our model, introducing interlayer tunneling results in avoided crossing of the bands due to hybridization of their X$_\text{A}$ and IX characters. The resonance of hybridization with equal composition of the two characters occurs at the position where the two decoupled bands cross. Away from this position, the degree of hybridization gradually reduces and the bands recover their original character. As shown in Fig.~\ref{mbs}, the position of the band crossing evolves with the twist angle due to the modification of moir\'e geometry, and at the same time, the energy of the IX band at the $\gamma$ point shifts. Since the optically active X$_\text{A}$ and IX hybridized states (X$_1$ and X$_2$) are located at the $\gamma$ point, their energies and oscillator strengths display twist-angle dependence.

The above evolution of exciton hybridization can be divided into several limits, which we elaborate for the R-type stacking. For the case of R-type: (i) When twist angle $\theta=0^\circ$, the decoupled IX and X$_\text{A}$ bands cross on the right side of the $\gamma$ point and the IX band is below X$_\text{A}$ at the $\gamma$ point. As a result, the lower energy hybridized exciton state X$_1$ has more IX character and the higher energy state X$_2$ has more X$_\text{A}$ character, implying a smaller oscillator strength of X$_1$ as compared to X$_2$. (ii) When $\theta$ increases, the position of the band crossing in momentum space moves towards the $\gamma$ point and the hybridization becomes stronger, transferring the oscillator strength from X$_2$ to X$_1$ state and pushing the X$_2$ state to higher energy. Note that because the decoupled IX band moves up at the $\gamma$ point, the X$_1$ state is also blue-shifted. (iii) When $\theta$ reaches $4.6^\circ$, the band crossing takes place exactly at the $\gamma$ point, giving rise to maximal hybridization. The X$_1$ and X$_2$ states have equal IX and X$_\text{A}$ characters and thus the same oscillator strength at this twist angle. (iv) When $\theta$ continues to increase, the band crossing position moves away to the left side of the $\gamma$ point and the decoupled IX and X$_\text{A}$ bands reverse the energy ordering at this point. Consequently, X$_1$ becomes more in X$_\text{A}$ character, while X$_2$ has more IX character. The oscillator strength of X$_1$ (X$_2$) thus increases (decreases). Meanwhile, the energy of X$_2$ state increases as the decoupled IX band moves up at the $\gamma$ point. The energy of the X$_1$ state also increases because of the reduced degree of hybridization and less avoided crossing at the $\gamma$ point. (v) When $\theta\gtrsim8^\circ$, the band crossing position is too far away from the $\gamma$ point and the degree of hybridization becomes vanishingly small. As a result, X$_\text{A}$ and IX bands are fully uncoupled and only one exciton state that corresponds to the A-exciton of MoTe$_2$ is optically active (the pure IX state has negligible oscillator strength). In this twist angle range, the energy and oscillator strength of the single exciton state show almost no variations. For H-type, the trend is similar to R-type, but the hybridized excitons have higher energies as the spin-singlet IX and X$_\text{A}$ are energetically closer due to the flipped spin-ordering of IX bands. In addition, the starting degree of hybridization at $60^\circ$ is higher and the maximal hybridization takes place for a smaller deviation from ideal anti-parallel alignment at an angle of $57.2^\circ$.

\vspace{16pt}
\noindent \textbf{Supplementary Note $\mathbf{10}$: Density functional theory}
\phantomsection
\addcontentsline{toc}{section}{Note $\mathbf{10}$: Density functional theory}
\vspace{8pt}

The theoretical simulations of the band dispersion parameters and the $g$-factors were performed using density functional theory (DFT) on the level of the Perdew-Becke-Ernzerhof (PBE) approximation as implemented in the Quantum Espresso package~\cite{Giannozzi2009}. The core electrons were replaced by fully-relativistic normconserving pseudopotentials from the PseudoDojo library~\cite{vanSetten2018}, where the $s$ and $p$ semi-core electrons of Mo are included in the set of valence electrons. As a preliminary step, we optimized the atomic positions and the in-plane lattice constants of isolated MoSe$_2$ and MoTe$_2$ monolayers until the interatomic forces and the stress were below thresholds of 0.01~eV/\r{A} and 0.01~GPa, respectively, while keeping a vacuum layer of at least 15~\r{A} in the out-of-plane direction in order to minimize interactions between periodic images. We included semi-empirical van der Waals corrections from the PBE+D3 scheme~\cite{Grimme2011}, which yields excellent predictions of the in- and out-of-plane lattice constants of layered systems. Integrations in reciprocal space were performed on a Monkhorst-Pack grid of $12\times12\times1$ points in the Brillouin zone. In this step, we did not explicitly include spin-orbit interactions (SOI). Based on the obtained optimized structure, we then simulated the electronic band energies, the corresponding Kohn-Sham orbitals and the interband optical matrix elements, and the vacuum level energies of the monolayers on the PBE level, with full inclusion of spin-orbit interactions. From the obtained bandstructures, we then derived the effective masses and orbital angular momenta of the relevant valence and conduction bands~\cite{Deilmann2020}. For the orbital angular momenta, the sum over states was performed over 2000 valence and conduction bands. Using Eq.~2 in the main text, we obtained a $g$-factor of $-4.6$ and $-4.3$ for the A-exciton in monolayer MoTe$_2$ and MoSe$_2$, in excellent agreement with the experimental results (see Fig.~S1).

\begin{table}[!h]
    \begin{ruledtabular}
    \begin{tabular}{rrrrrrrr}       
                 & $m_c/m_0$ & $m_{c+1}/m_0$ & $m_v/m_0$ & $\Delta_\mathrm{SO}$ & $L_c$ & $L_{c+1}$ & $L_v$ \\  
        \hline
        MoTe$_2$ &      0.58 &          0.67 &   $-0.68$ &                         69 & 1.586 &     1.204 & 3.872 \\
        MoSe$_2$ &      0.55 &          0.63 &   $-0.64$ &                         23 & 1.798 &     1.526 & 3.977 \\
    \end{tabular}
    \caption{Material parameters at the $K$ points from DFT calculations. $m_c$ ($m_{c+1}$) is the effective electron mass of the first (second) conduction band with free electron mass $m_0$; $m_v$ is the effective mass of the highest valence band; $\Delta_\mathrm{SO}$ is the spin-orbit splitting in the conduction band, in units of meV. $L_c$ and $L_{c+1}$ are the orbital angular momenta of the first and second conduction bands, and $L_v$ is the orbital angular momentum of the highest valence band.}
    \label{parameters}
    \end{ruledtabular}
\end{table}

%